\documentclass[usegraphicx]{mn2e}
\begin{document}
\title[Radiation reaction and the pitch angle changes]
{Radiation reaction and the pitch angle changes for a charge undergoing synchrotron losses}
\author[Singal]{Ashok K. Singal\\
{Astronomy and Astrophysics Division, Physical Research Laboratory, 
Navrangpura, Ahmedabad - 380 009, India}\\
\\E-mail: asingal@prl.res.in}

\date{Accepted . Received ; in original form }
\maketitle

\begin{abstract}
In the derivation of synchrotron radiation formulas it has been assumed that the pitch angle of a 
charge remains constant during the radiation process. However from the radiation reaction formula, while 
the component of the velocity vector 
perpendicular to the magnetic field reduces in magnitude due to radiative losses, the  
parallel component does not undergo any change during radiation. Therefore there is a change in the ratio of 
the two components, implying a change in the pitch angle. 
We derive the exact formula for the change in energy of radiating electrons by taking into account the 
change of the pitch angle due to radiative losses. From this we derive the characteristic decay time of synchrotron electrons over 
which they turn from highly relativistic into mildly relativistic ones. 
\end{abstract} 
\section{INTRODUCTION}
Formulas for synchrotron radiative losses were derived more than about 
50 years back and have been in use ever since for calculating radiative 
losses in a variety of radio sources. Since the radiation is beamed with its angular 
distribution confined to a narrow cone around the direction of the instantaneous velocity vector of the 
charge, the momentum loss of the radiating charge is expected to be along the direction 
of motion of the charge. Therefore though the kinetic energy of the radiating charge would be reducing due to 
the radiation losses, its direction of motion should not undergo any changes and there ought not be any changes in the 
pitch angle of the charge during the radiation process. Thus in these 
formulas, it has always been assumed that the pitch angle of the radiating 
charged particle remains constant and the 
dynamics and the life-time of radiating electrons are accordingly derived (Kardashev 1962). 
This formulation is now a standard text-book material (Pacholczyk 1970; Rybicki \& Lightman 1979).

In this  formulation, the power loss rate is written as (Kardashev 1962; Pacholczyk 1970; Rybicki \& Lightman 1979),
\begin{equation}
\label{eq:30.1}
\frac{{\rm d}{\cal E}}{{\rm d}t} = -\zeta \sin^2 \theta\; {\cal E}^2\;,
\end{equation}
where
\begin{eqnarray}
\label{eq:30.1a}
\zeta = \frac{2e^4\, B^2} {3m_{o}^4c^7}=2.37 \times 10^{-3}B^2\;  {\rm erg}^{-1} {\rm s}^{-1}\;.
\end{eqnarray}

Let ${\cal E}_{\rm o}$ be the initial energy  at $t=0$, then from Eq.~(\ref{eq:30.1}) the energy of the radiating charge at $t=\tau$ is 
calculated to be,
\begin{equation}
\label{eq:30.4}
{\cal E}= \frac{{\cal E}_{\rm o}}{1+\zeta\sin^2 \theta\;  \tau\; {\cal E}_{\rm o}}\;.
\end{equation}
Here pitch angle $\theta$ is treated as a constant of motion. 

From Eq.~(\ref{eq:30.4}), radiative losses have been worked out to calculate the life-time of the radiating electrons
(Kardashev 1962; Kellermann 1964; van der Laan \& Perola 1969; Pacholczyk 1970; Miley 1980). 
For instance, from Eq.~(\ref{eq:30.4}) it follows that the electron loses half of its energy in a time 
$\tau_{1/2} = 1/(\zeta\sin^2 \theta \;{\cal E}_{\rm o})$,  
and the synchrotron source that has been radiating for a time, 
say $\tau$, will have no electrons above an energy ${\cal E}_\tau=1/(\zeta\sin^2 \theta\;\tau)$, 
even if we started with electrons 
with an infinite energy to begin with. This in turn also implies  that  there will  be a 
break  in the spectrum at a  frequency 
$\nu_\tau \propto \tau^{-2}$ for such a source. Such observed breaks in the spectra have been used to estimate 
the ages of extragalactic radio sources (van der Laan \& Perola 1969) and supernovae remnants (Chandra et al. 2004). 
Evolution in the spectra of such radio sources have been studied for different scenarios of fresh particle injections 
(Kardashev 1962; Pacholczyk 1970). These formulas have appeared in a number of  books and review articles, 
and a large number of papers in the literature have made use of these formulas. 

Actually it is not exactly true that the pitch angle remains constant in spite of radiation losses. But it is still a very good 
approximation for highly relativistic particles, like synchrortron electrons in radio galaxies. However, as it will be shown,   
this approximation does not remain valid in trans- and sub-relativistic regimes. From radiation reaction it has been 
shown (Petrosian 1985) that in the case of synchrotron losses while 
the component of the velocity vector perpendicular to the magnetic field reduces in magnitude, the  
parallel component does not undergo any change due to radiation. Thus a change in the ratio of 
the two components takes place, implying a change in the pitch angle. Therefore one must consider this change in pitch angle 
while calculating radiation losses and the consequent life times of synchrotron electrons.
Moreover, Eq.~(\ref{eq:30.1}) is an approximate expression, again valid only in highly relativistic cases. 
The exact equation for radiative energy loss rate is (Melrose 1971; Jackson 1975; Longair 2011), 
\begin{eqnarray}
\label{eq:30.1b}
\frac{{\rm d}{\cal E}}{{\rm d}t} = -\zeta \beta ^2\sin^2 \theta\;{\cal E}^2,
\end{eqnarray}
which also changes the solution in Eq.~(\ref{eq:30.4}). We shall derive the exact formulation for radiative losses, 
applicable even to non-relativistic or mildly relativistic cases, simultaneously taking into account the pitch angle changes. 
\section{RADIATION REACTION}
Radiation reaction was first derived by Lorentz (1909) from the self-force of a small charged sphere, 
and the detailed derivation is available in various forms in many text-books (Schott, 1912; Jackson, 1975; 
Panofsky \& Phillips 1962; Heitler 1954; Griffiths 1999; Yaghjian 2006). But recently it has been 
derived even for a ``point'' charge by examining the electromagnetic fields in the neighbourhood of the point charge and then 
from momentum conservation employing Maxwell-stress tensor (Singal 2015a,b). 
Of course the two independent methods do lead to the same result, 
\begin{equation}
\label{eq:1d6}
{\bf {f}}=\frac{2e^{2}}{3 c^{2}}\ddot{\mbox{\boldmath $\beta$}}\;.
\end{equation}
This is the self-force on the charge in its instantaneous rest frame. 

The force due to the radiation reaction on a charge with relativistic motion (a generalization of Eq.~(\ref{eq:1d6})) 
for a synchrotron case (i.e., where $\dot{\mbox{\boldmath $\beta$}}\cdot{\mbox{\boldmath $\beta$}}=0$) is given by 
(Heitler 1954; Singal 2016),
\begin{equation}
\label{eq:10.1}
{\bf F}=\frac{2e^{2}}{3c^{2}}\gamma ^{4}\left[\ddot{\mbox{\boldmath $\beta$}}+ {\mbox{\boldmath $\beta$}}\times({\mbox{\boldmath $\beta$}}\times\ddot{\mbox{\boldmath $\beta$}})\right]\;,
\end{equation}
where 
\begin{equation}
\label{eq:10.1.1}
\ddot{\mbox{\boldmath $\beta$}}=\frac{e}{m_{\rm o}\gamma c}\dot{\mbox{\boldmath $\beta$}}\times{\bf B}
=-\frac{e^2}{m_{\rm o}^2\gamma^2 c^2}B^2{\mbox{\boldmath $\beta$}}_\perp\;.
\end{equation}
We then get force components parallel and perpendicular to the magnetic field direction (Petrosian 1985) as, 
\begin{equation}
\label{eq:10.1.2}
{F}_\parallel=\frac{2e^2}{3c^2}\gamma ^{4} \ddot{\beta} \beta_\parallel {\beta}_\perp
=-\frac{2e^4}{3m_{\rm o}^2c^4}\gamma ^{2} B^2 {\beta}_\perp^2 \beta_\parallel \;,
\end{equation}
\begin{equation}
\label{eq:10.1.3}
{F}_\perp=\frac{2e^{2}}{3c^{2}}\frac{\gamma ^{4}\ddot{\beta}}{\gamma_\parallel^2}
=-\frac{2e^4}{3m_{\rm o}^2c^4}\frac{\gamma ^{2 }B^2 {\beta}_\perp}{\gamma_\parallel^2}\;,
\end{equation}
where $\gamma_\parallel=\surd(1- \beta_\parallel^2)$. 

The negative sign indicates that the force component in each case is in a direction opposite to the direction of the 
corresponding velocity component of the charge. However, the ratio of the force components is not equal to that of the velocity components, 
\begin{equation}
\label{eq:10.1.3a}
{F}_\perp/{F}_\parallel=1/(\gamma_\parallel^2 \beta_\parallel \beta_\perp)\ne  \beta_\perp / \beta_\parallel\;.
\end{equation}
which implies that the force vector is not parallel to the velocity vector. 
Also, in spite of a finite ${F}_\parallel$, there is no acceleration $\dot{\beta}_\parallel$ along the $z$-axis. 
It can be recalled that in special relativity, force and acceleration vectors are not
always parallel, e.g., in a case where force is not parallel to the velocity vector, the acceleration need not be 
along the direction of the force. When the applied force is either parallel to or perpendicular to the velocity vector, it is 
only then that the acceleration is along the direction of force (see e.g., Tolman 1934). 
It has to be further kept in mind that the acceleration we are talking 
about here is not that due to the force by the magnetic field on the moving charge (which is perpendicular to the 
instantaneously velocity of the charge), but the acceleration (or rather a deceleration) caused on the charge due to the 
radiation reaction force. 

\section{RADIATIVE LIFE TIMES}
The radiation drag will reduce only the perpendicular velocity component, ${\beta}_\perp$, of the charge 
(i.e., $\dot{\beta}_\parallel=0$). The rate of change of the parallel component of the momentum is,
\begin{eqnarray}
\label{eq:10.1.4}
\frac{{\rm d}(m_{\rm o} \gamma\beta_\parallel c)}{{\rm d}t}=m_{\rm o} c \frac{{\rm d}{\gamma}}{{\rm d}t} \beta_\parallel 
=m_{\rm o} c\gamma^3 \dot{\beta}_\perp {\beta}_\perp \beta_\parallel\;.
\end{eqnarray}
Equating it to $F_\parallel$ (Eq.~(\ref{eq:10.1.2})), the force due to radiation reaction, we can write,
\begin{eqnarray}
\label{eq:10.1.5}
m_{\rm o} c\gamma^3 \dot{\beta}_\perp {\beta}_\perp \beta_\parallel
=-\frac{2e^4}{3m_{\rm o}^2c^4}\gamma ^{2} B^2 {\beta}_\perp^2 \beta_\parallel\;,
\end{eqnarray}
which simplifies to,
\begin{equation}
\label{eq:30.6c}
{\dot{\beta}_\perp} =  \frac{-{\eta}{\beta_\perp}}{\gamma}\;.
\end{equation}
Here ${\eta}=2e^4 B^2/(3m_{\rm o}^3c^5)=1.94 \times 10^{-9}B^2\; {\rm s}^{-1}$.
On the other hand, from the perpendicular component of the rate of change of momentum (Eq.~(\ref{eq:10.1.3})), we have,
\begin{eqnarray}
\label{eq:10.1.6}
\frac{{\rm d}{\gamma}}{{\rm d}t} {\beta}_\perp + \gamma \dot{\beta}_\perp 
=-{\eta}\frac{\gamma ^{2 }{\beta}_\perp}{\gamma_\parallel^2}\;,
\end{eqnarray}
Eliminating $\dot{\beta}_\perp$ with the help of Eq.~(\ref{eq:30.6c}) we get (also see, Petrosian 1985),
\begin{equation}
\label{eq:30.2.1}
\frac{{\rm d}{\gamma}}{{\rm d}t} =  -{\eta}\left(\frac{\gamma^2}{\gamma_\parallel^2}-1\right)\;.
\end{equation}
 
Equation~(\ref{eq:30.2.1}) has a solution,
\begin{equation}
\label{eq:30.2c}
\tanh^{-1}\frac{\gamma_\parallel}{\gamma}= \frac{\eta t}{\gamma_\parallel} + a\;.
\end{equation}
where $a$ is a constant of integration. Let ${\gamma_{\rm o}}$ correspond to the initial energy of the charge at $t=0$, implying 
${\gamma_\parallel/\gamma_{\rm o}}=\tanh a$, then at $t=\tau$ we have,
\begin{equation}
\label{eq:30.2e1}
\tanh^{-1}\frac{\gamma_\parallel}{\gamma} = \tanh^{-1}\frac{\gamma_\parallel}{{\gamma_{\rm o}}}+\frac{{\eta} \tau}{\gamma_\parallel}\;.
\end{equation}
This is a general solution for all values of $\gamma$. 
We can rewrite it as,
\begin{equation}
\label{eq:30.2e2}
\gamma = \gamma_\parallel\; \frac{{\gamma_{\rm o}}+\gamma_\parallel\tanh({\eta} \tau/\gamma_\parallel)}
{{\gamma_{\rm o}}\tanh({\eta} \tau/\gamma_\parallel)+\gamma_\parallel}\;.
\end{equation}
For an initially ultra-relativistic charge ($\gamma_{\rm o}\gg1, \beta_{\rm o} \approx 1$), we have 
$1/\gamma_\parallel =\surd(1-\beta_{\rm o}^{2}\cos^2 \theta_{\rm o})\approx \sin\theta_{\rm o}$. That also implies that 
(except for initially small pitch angle cases) $\sin \theta_{\rm o} \gg 1/\gamma_{\rm o}$ or $\gamma_\parallel / \gamma_{\rm o} \ll 1$, and from 
Eq.~(\ref{eq:30.2e1}) we could write,
\begin{equation}
\label{eq:30.2e1a}
\tanh^{-1}\frac{1}{\gamma \sin\theta_{\rm o}} = {{\eta} \tau}{\sin\theta_{\rm o}}\;.
\end{equation}
This implies that for $\tau = \gamma_\parallel/ \eta \approx 1/ (\eta \sin \theta_{\rm o})$, we have $\gamma \approx 1.3/\sin \theta_{\rm o}$. 
Thus even if an electron had started with an almost infinite energy, it loses most of its kinetic energy in a time interval of the order 
of $1/ \eta$, reducing to perhaps a mildly relativistic status (for not too small an initial pitch angle). 
For instance let us consider $\gamma_{\rm o}=10^3$ and $\theta_{\rm o}=\pi/4$, then $\gamma_\parallel \approx 1/\sin \theta_{\rm o}=\surd{2}$, then from 
Eq.~(\ref{eq:30.2e1}) or Eq.~(\ref{eq:30.2e1a}) we get for $\tau = 1/\eta$, $\gamma=2.3$. In another example, taking $\gamma_{\rm o}=10^4$ and 
$\theta_{\rm o}=\pi/3$,  for $\tau = 1/\eta$ we get $\gamma_\parallel \approx 1/\sin \theta_{\rm o}=2/\surd{3}$ and $\gamma=1.7$.
Thus  $1/ \eta$ represents the characteristic decay time of synchrotron electrons over 
which they turn from ultra-relativistic into mildly relativistic ones. 

On a shorter time scale where $\gamma$ is still highly relativistic, we could write
\begin{equation}
\label{eq:30.2f}
\frac{1}{\gamma} = \frac{1}{{\gamma_{\rm o}}}+\frac{{\eta} \tau}{\gamma_\parallel^2}\;,
\end{equation}
Then we have,
\begin{equation}
\label{eq:30.2g}
\gamma = \frac{\gamma_{\rm o}}{1+{\eta}\tau \gamma_{\rm o} / \gamma^2_\parallel}
=\frac{{\gamma_{\rm o}}}{1+{\eta}\tau\gamma_{\rm o}\sin^2 \theta_{\rm o}}\;.
\end{equation}
Equation~(\ref{eq:30.2g}) is the same relation as in Eq.~(\ref{eq:30.4}), which tells us that 
there is an upper energy-cutoff for highly relativistic electrons in a synchrotron source, 
$\gamma_{\tau}=\gamma^2_\parallel/({\eta} \tau)=1/({\eta} \tau \sin^2 \theta_{\rm o})$. Also, while $\tau_{1/2} 
= 1/(\eta\sin^2 \theta_{\rm o} \;{\gamma}_{\rm o})$ gives half-life time of the synchrotron electrons (Kardashev 1962;  
Pacholczyk 1970; Rybicki \& Lightman 1979), $\tau \sim 1/\eta$ gives decay time over which radiating synchrotron electrons 
turn from ultra-relativistic into mildly relativistic ones. 
This followed from Eq.~(\ref{eq:30.2e1}) which gives an exact formula, true for all $\gamma$, including mildly relativistic electrons and 
implicitly includes the effects of the pitch angle changes. An explicit expression for the pitch angle changes is derived in the next section.
\section{PITCH ANGLE CHANGES}
Both $\beta$ and $\theta$ in $\beta_\perp=\beta \sin \theta$ are functions of time. Therefore we can rewrite Eq.~(\ref{eq:30.6c}) as,
\begin{equation}
\label{eq:30.6c1}
{\beta \cos \theta \frac{{\rm d}\theta}{{\rm d}t} + \dot{\beta} \sin \theta} =  \frac{-{\eta}\beta \sin \theta} {\gamma}\;.
\end{equation}
Also since $\dot{\beta}_\parallel=0$, we have,
\begin{equation}
\label{eq:30.6d}
\beta \sin \theta \frac{{\rm d}\theta}{{\rm d}t} = \dot{\beta} \cos \theta\;.
\end{equation}
Eliminating $\dot{\beta}$ from Eqs.~(\ref{eq:30.6c1}) and (\ref{eq:30.6d}), we get (also see, Petrosian 1985),
\begin{equation}
\label{eq:30.6e}
\frac{{\rm d}\theta}{{\rm d}t}= \frac{-{\eta}\sin \theta \cos \theta}{\gamma}=\frac{-{\eta}\sin 2\theta}{2\gamma}\;.
\end{equation}
This is the relation for the rate of change of the pitch angle of a charge undergoing synchrotron radiative losses. 
The negative sign implies that the pitch angle decreases with time and aligns with the magnetic field. The rate of alignment 
is very slow for low pitch angles ($\theta \approx 0$) as well as for high pitch angles ($\theta \approx \pi/2$), and the highest 
rate of change of the pitch angle is for $\theta=\pi/4$.

With the help of Eq.~(\ref{eq:30.2c}), we can integrate Eq.~(\ref{eq:30.6e}) to get,
\begin{equation}
\label{eq:30.6i}
{\tan \theta} = \frac{{\tan \theta_{\rm o}}}{\cosh \left(\frac{{\eta} \tau}{\gamma_\parallel}\right)
+ \frac{\gamma_\parallel}{\gamma_{\rm o}} \sinh \left(\frac{{\eta} \tau}{\gamma_\parallel} \right)}\;.
\end{equation}

In Eq.~(\ref{eq:30.6i}), $\theta < \theta_{\rm o}$, because pitch angle always reduces with time. There are many notable points. If 
$\theta_{\rm o} =\pi/2$, then $\theta =\pi/2$ also, which is because if the pitch angle is $\pi/2$, then the radiating electron always moves in 
a circular path in the plane perpendicular to the magnetic field. And if $\theta_{\rm o} =0$, then $\theta=0$ too as there is no more reduction 
in the pitch angle. For any $0<\theta_{\rm o} <\pi/2$, $\theta \rightarrow 0$ as $\tau \rightarrow \infty$. For large $\gamma_{\rm o}$ values, 
\begin{equation}
\label{eq:30.6j}
{\tan \theta} = \frac{{\tan \theta_{\rm o}}}{\cosh \left({{\eta} \tau}{\sin \theta_{\rm o}}\right)}\;,
\end{equation}
which can be used to estimate change in pitch angle with time. For example for say, $\theta_{\rm o} =\pi/3$, and $\theta =\pi/6$, 
$\cosh ({\eta} \tau \sin \theta_{\rm o})=3$, which gives $\tau \approx 2 /\eta$ for this change in the pitch angle.
Thus there are appreciable pitch angle changes in time $\tau \sim 1/\eta$ (except for 
in the vicinity of very small pitch angles). 

All charges of a given energy and pitch angles, directed towards the observer in a narrow angle $1/\gamma$ around the line of sight 
not only lose energy but will even get shifted outside the angle $1/\gamma$ around the line of sight towards the observer, in a time 
$\tau \sim 1/\eta$. Thus in a mono-energetic and a narrow pitch angle distribution, the pitch angle changes might be quite relevant. 
But it may be of less importance when there is a wide angular (isotropic!) distribution of pitch angles.

In case of compact radio sources with ${\bf B} \sim 10^{-3}$G the time periods for pitch angle alignments may be $\sim 10^{8}$ years, 
but in extended radio sources where ${\bf B} \sim 10^{-5}$G, the pitch angle alignment would be much slower. 
However in mildly relativistic cases (see e.g. Petrosian 1981) pitch angle changes may be of sufficient 
importance and might not be ignored. Also in cases where 
magnetic field is much stronger, e.g., in case of curvature radiation models in pulsars, where magnetic fields could vary from 
${\bf B} \sim 10^{12}$G near the neutron star surface to $\sim 10^{6}$G in the outer magnetosphere, the pitch angle changes will be 
rather fast and that is why the electrons will stream along the curved magnetic field lines, giving rise to curvature radiation. 
It may be mentioned that our discussion is well within the classical synchrotron radiation regime where energy loss time scales are 
much larger than the gyro period and the quantum effects do not become applicable (see e.g., Brainerd 1987;  Brainerd \& Petrosian 1987).

\section{CONCLUSIONS}
We showed that the argument used in the standard formulation of the synchrotron radition that the radiation is beamed along the 
direction of motion and therefore radiation losses should not cause any change in the pitch angle of the charge is not correct   
and that the pitch angle in general varies. It is because  from the radiation reaction formula while the component of the velocity vector 
perpendicular to the magnetic field reduces in magnitude due to radiative losses, the  
parallel component does not undergo any change during radiation, which implies a change in the pitch angle. 
We derived the exact formula 
for the change in energy of radiating electrons by taking into account the 
changes in the pitch angle due to radiative losses. This way we derived the characteristic decay time of synchrotron electrons over 
which they turn from highly relativistic into mildly relativistic ones. 
Further, an explicit formula for the pitch angle changes with time was also derived.

\end{document}